\documentclass[runningheads]{llncs}
\usepackage[T1]{fontenc}

\usepackage{graphicx}

\usepackage{cite}
\usepackage{amsmath,amssymb,amsfonts}
\usepackage{graphicx}
\usepackage{textcomp}
\usepackage{xcolor}
\usepackage{lipsum}% generate text for the example
\usepackage{algorithm} % used for the other package
\usepackage{algpseudocode}
\usepackage{xspace}
\usepackage{todonotes}
\usepackage{hyperref}
\usepackage{soul}
\usepackage[shortlabels]{enumitem}

\newcommand{\coolname}{$\mathtt{RADD}$\xspace}

\begin{document}
\title{Runtime Anomaly Detection for Drones:\\ An Integrated Rule-Mining and Unsupervised-Learning Approach}
\titlerunning{Runtime Anomaly Detection for Drones: An Integrated Approach}

\author{Ivan Tan \and Wei Minn \and Christopher M. Poskitt \and Lwin Khin Shar \and Lingxiao~Jiang}
\authorrunning{I. Tan et al.}

\institute{Singapore Management University, Singapore\\
\email{\{ivantan,wei.minn.2023,cposkitt,lkshar,lxjiang\}@smu.edu.sg}}
\maketitle             
\begin{abstract}
    Unmanned Aerial Vehicles (UAVs), commonly referred to as drones, have witnessed a remarkable surge in popularity due to their versatile applications.
    These cyber-physical systems depend on multiple sensor inputs, such as cameras, GPS receivers, accelerometers, and gyroscopes, with faults potentially leading to physical instability and serious safety concerns.
    To mitigate such risks, anomaly detection has emerged as a crucial safeguarding mechanism, capable of identifying the physical manifestations of emerging issues and allowing operators to take preemptive action at runtime.
    Recent anomaly detection methods based on LSTM neural networks have shown promising results, but three challenges persist: the need for models that can generalise across the diverse mission profiles of drones; the need for interpretability, enabling operators to understand the nature of detected problems; and the need for capturing domain knowledge that is difficult to infer solely from log data.
    Motivated by these challenges, this paper introduces \coolname, an integrated approach to anomaly detection in drones that combines rule mining and unsupervised learning.
    In particular, we leverage rules (or invariants) to capture expected relationships between sensors and actuators during missions, and utilise unsupervised learning techniques to cover more subtle relationships that the rules may have missed.
    We implement this approach using the ArduPilot drone software in the Gazebo simulator, utilising 44 rules derived across the main phases of drone missions, in conjunction with an ensemble of five unsupervised learning models.
    We find that our integrated approach successfully detects 93.84\%\ of anomalies over six types of faults with a low false positive rate (2.33\%), and can be deployed effectively at runtime.
    Furthermore, \coolname outperforms a state-of-the-art LSTM-based method in detecting the different types of faults evaluated in our study.
\end{abstract}


\section{Introduction}
\label{sec:intro}

In recent years, Unmanned Aerial Vehicles~(UAVs), commonly known as drones, have surged in popularity owing to their use in areas as diverse as aerial photography, construction inspection, and product delivery.
These sophisticated and complex cyber-physical systems~(CPSs) are operated remotely with the help of inputs from multiple sensors, including cameras, GPS receivers, accelerometers, and gyroscopes, all of which the control software uses to determine appropriate motor actuations.
Failures in sensors and actuators (e.g.~due to intentional sound noise attacks~\cite{son2015rocking}), and even environmental factors (e.g.~strong wind~\cite{bugs2}) can impact the stability of a drone's operation, which can lead to serious and sometimes life-threatening hazards.
Detecting and diagnosing these faults accurately and promptly is thus crucial to ensuring the safety and success of the drone and its mission.

One approach for achieving this is \emph{anomaly detection}, a commonly used technique for identifying behaviours that deviate significantly from the norm.
In this context, behaviours are inferred from logs of timestamped sensor readings and actuator states, with anomalies defined as abnormal data points or sequences within these logs.
Anomaly detection systems for CPSs can be built in a number of different ways, but a particularly popular approach is to train a machine learning~(ML) model on historical data from the system.
If labelled data is available, then supervised learning approaches like SVM can be used to train a model that classifies new data points as normal or anomalous~\cite{Chen-Poskitt-Sun18a}. However, in drone anomaly detection, labelled data is often unavailable due to the volatile nature of anomalies and the substantial manual effort required for labelling, even when anomalies are known. In the absence of labelled data, unsupervised approaches like clustering~\cite{Kiss-et_al15a}, local outlier factor~\cite{7925416}, and one-class SVM~\cite{Inoue-et_al17a} can be used to learn a decision boundary that encapsulates the normal data points.
Deep learning approaches, in which neural networks learn the complex non-linear patterns in data sets, can also be used for anomaly detection by serving as time series predictors, with the anomaly alarm raised when predicted physical states differ significantly from the ones eventually observed~\cite{Goh-et_al17a,Kravchik-Shabtai18a,Kravchik-Shabtai22a}.
Anomalies can also be detected by evaluating data points against rules (or invariants), in which expected relationships between sensors and actuators are extracted from data sets~\cite{Feng-et_al19a}, control programs and the laws of physics~\cite{Choi-et_al18a}, or alternatively, are defined based on \emph{a priori} knowledge~\cite{Adepu-Mathur16a,Yoong-et_al21a}.

DronLomaly~\cite{Shar-et_al22a} has taken some first steps towards applying anomaly detection techniques to drones, specifically, by utilising LSTM netural networks to predict future sensor/actuator states and marking anomalies when the actually observed states diverge.
Despite some promising results for three sets of flight logs, there remain a number of challenges that are particularly pertinent to drones.
First is the risk of overfitting to the training data from sample flying logs.
Models must be able to \emph{generalise}, as no two missions are the same: flying in an open field is different to flying in a crowded city, not to mention the differences that arise due to weather.
Second, some existing methods, such as LSTM-based ones, lack \emph{interpretability}, making it difficult for operators to understand the nature of an anomaly and to decide what the appropriate action is to take.
Finally, drones are often required to satisfy properties that are simple to define with \emph{domain knowledge} but may not always be easily inferred from log data.
For example, regulations concerning speed and height restrictions, or acceptable lag between an operator issuing a command and the drone enacting it.
Therefore, we are motivated to explore alternative anomaly detection approaches that overcome these specific challenges for drones.

In this paper, we propose \coolname~\emph{(Runtime Anomaly Detection for Drones)}, an integrated approach to anomaly detection for drones that combines rule-checking and unsupervised-based methods.
First, we leverage rules (or invariants) to capture expected relationships between sensors and actuators throughout the main phases of drone missions.
Our rule sets are automatically mined from log data using the Apriori algorithm, then complemented with additional custom rules based on domain knowledge (e.g.~acceptable latency).
Second, we utilise an ensemble of unsupervised models in order to be able to identify more subtle deviations that the rules may have missed.
If a rule is violated or (a majority of) the ensemble reports an anomaly, then an anomaly is reported.
By eschewing deep neural networks for an integrated rule- and unsupervised-based approach, \coolname identifies anomalies across a wide range of missions and reports them alongside the human-readable rules used for their detection, making the results more understandable to users than in previous approaches.

We implemented \coolname for the ArduPilot drone software~\cite{ardupilot} in the Gazebo simulator~\cite{SITL}, utilising 44 rules mined across the five main phases of generic drone missions, as well as an ensemble of five unsupervised models (k-means, DBSCAN, OPTICS, LOF, and SVM) that operate by majority vote.
We evaluated the approach against multiple datasets covering normal flights as well as multiple faults, such as engine failures, sensor failures, and heavy wind.
We found that our integrated approach was able to detect an average of $93.84\%$ of anomalies over six types of faults with a low rate of false positives.
\coolname's detection rate rises to over $99\%$ of anomalies for the specific cases of heavy wind and various sensor faults.
Furthermore, we found that \coolname outperformed the baseline of DronLomaly's LSTM model~\cite{Shar-et_al22a}, and in an ablation study, observed that our approach requires \emph{both} the rule- and unsupervised-based components to maintain its high accuracy across missions in general.
Finally, we observed that \coolname was efficient enough to be deployed at runtime and detect anomalies during missions.

The main contributions of our work include:

\begin{itemize}
    \item The first anomaly detection technique for drones based on the integration of rule-mining and unsupervised learning.%, and the introduction of phases.
    \item An implementation for ArduPilot, including 44 rules defined over several mission phases, and an ensemble of five unsupervised models.
    \item An experimental evaluation in which $93.84\%$ of anomalies over six types of faults (rising to over $99\%$ for heavy wind and sensor faults) with a false positive rate of 2.33\%.
    \item An interpretability study evaluating \coolname, where it achieved 6.6 out of 7 on a Likert scale rating as compared to raw data (3.2) in terms of understanding anomalies identified. 
    \item A demonstration of the feasibility of verifying the absence of drone anomalies at runtime using \coolname.
    \item Source code and datasets used for RADD, available at \cite{RADDnew}.
\end{itemize}

\noindent \coolname's integration of rules and unsupervised models addresses the challenges of generalising across missions and capturing domain knowledge, while ensuring that detected anomalies are accompanied by human-readable rules—enhancing interpretability compared to state-of-the-art LSTM methods~\cite{Shar-et_al22a}.

The rest of the paper is organised as follows.
Section~\ref{sec:background} provides background on drones and the components of the simulation environment we used.
Section~\ref{sec:methodology} presents our approach.
Section~\ref{sec:experiment} presents our research questions, experiments, and results.
Section~\ref{sec:related-work} discusses the related works on anomaly detection techniques for drones and other cyber-physical systems.
%Section~\ref{sec:evaluation} presents the results of our experiment.
Finally, Section~\ref{sec:conclusion} concludes the paper and discusses future work.

\section{Background}
\label{sec:background}

In this section, we present some background on the structure of a typical drone system, its logs, and our simulation environment.

\begin{figure}[t]
    \centering
    \includegraphics[width=0.7\linewidth]{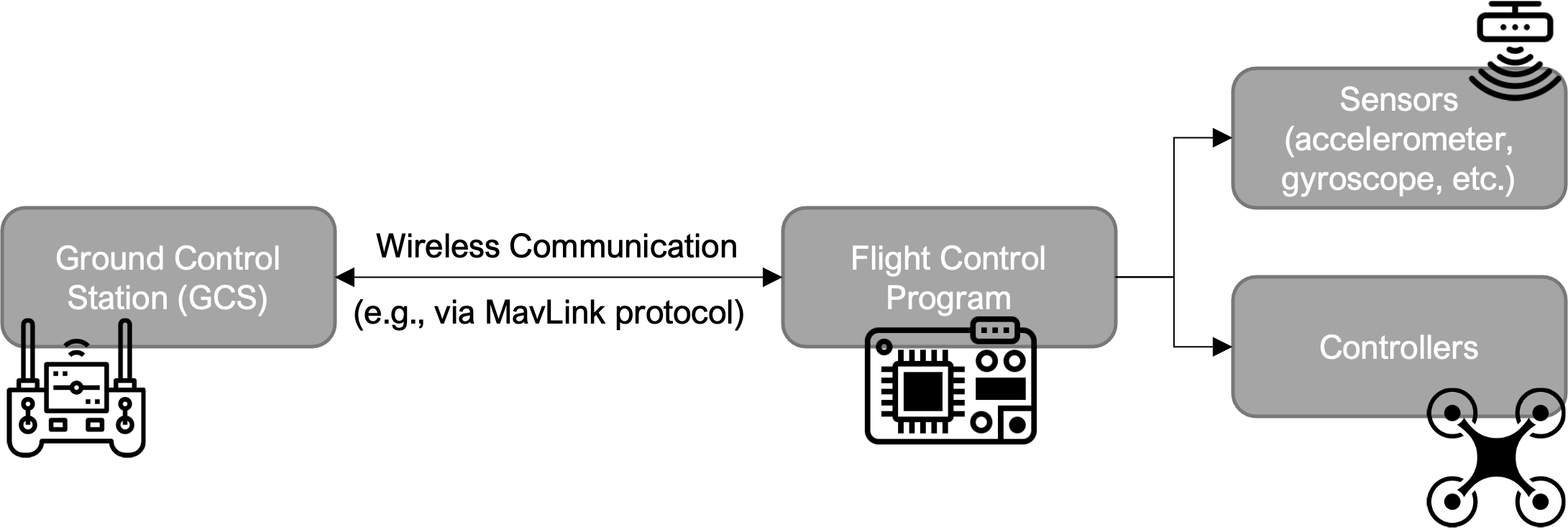}
    \caption{Overview of a typical drone system}
    \label{fig:drone-system}
\end{figure}

\noindent
\textbf{Drone Setup and Logs.} Typically, a drone system consists of a ground control station~(GCS), flight control program, sensors, and controllers. Figure~\ref{fig:drone-system} shows a block diagram of a drone system. 

The flight control program can be embedded in a small computing device such as a Raspberry Pi. It is the heart of the system that controls the drone's movements and operations. It may operate autonomously through an autopilot function for accomplishing a mission (e.g.~flying along a preset flight path with multiple waypoints) or by dynamic commands sent from the GCS via a communication protocol, e.g.~via a MavLink (Micro Air Vehicle Link) message. It reads data from sensors, such as the Global Positioning System (GPS), barometers, accelerometers, gyroscopes, and thermometers, to detect the physical states of the drone, and sends actuator commands to the physical controllers according to the pre-programmed flight mission or the dynamic commands sent by the user via the GCS.
It may control multiple physical controllers, e.g.~the z-axis controller for movement along vertical direction, the x-axis and y-axis controllers for movement along horizontal directions, and roll, pitch, and yaw controllers for rotation. 

Drones are subjected to extreme conditions including vibration, noise, and environmental stresses. Ideally, sensors/actuators in drones should have high shock survival capability and be fast enough to capture/withstand vibrations, and sensors should be able to filter noise such as GPS signal interference. The performance of sensors and actuators should not vary with changes in environment parameters such as temperature and humidity. When this is not the case, failures or anomalies are said to occur. It is important to detect them in real time so that an operator can take a corrective action promptly.

A majority of drones, especially industrial drones, log flight data during missions. This data typically includes the flight status, sensor readings, actuator outputs, and other information such as configuration parameter values. Behaviours of the drone including anomalies can be inferred from such logs.
These logs can be accessed by a program embedded in the drone or in the GCS during runtime\footnote{As a proof-of-concept and for our experiments, we developed a C/C++ program for DJI drones and embedded it in a Raspberry Pi device attached to a DJI drone.}.

\noindent
\textbf{Simulation Environment.}
\label{sec:sim}
To simulate the drones' missions and gather the log data, we used a combination of Ardupilot, Pymavlink, and Gazebo. Ardupilot is an open-source control program for drones and other unmanned vehicles that provides a range of features for controlling the flight of the vehicle, including waypoint navigation, automated takeoff and landing, stabilisation, and various flight modes.
Pymavlink is a Python implementation of the MavLink communication protocol, which is a lightweight messaging protocol designed for communication between unmanned vehicles and their ground control stations. The protocol provides a set of Python modules for encoding and decoding these MAVLink messages, as well as tools for working with MavLink logs and performing simulations. 
The Gazebo Simulator is an open-source 3D robotics simulation software widely used for testing and developing robotics algorithms and systems. It provides a high-fidelity physics engine that simulates the dynamics and kinematics of robots and other objects in a virtual environment. 

\section{Our Approach}
\label{sec:methodology}

In this section, we present \coolname \emph{(Runtime Anomaly Detection for Drones)}, a new approach that integrates rule mining and unsupervised learning techniques to detect and interpret anomalies across different drone mission profiles.
First, we propose segmenting generic drone missions into several key phases, which is essential, as rule associations may differ across the phases.
Second, we describe our rule mining approach for drones, based on the Apriori algorithm.
These rules are complemented with additional custom rules based on domain knowledge (e.g.~acceptable latency between issuing a command and the drone acting on it).
Finally, we propose an ensemble of five unsupervised models to cover the gaps missed by our rules, and explain how the rules and models can be used together to check anomalies at runtime.

\noindent
\textbf{Deriving the Mission Phases.} Segmenting missions into general phases is important for rule mining, because certain relationships between values in the data logs only hold during certain parts of the mission.

For the case of ArduPilot, throttle while the drone is initialising (i.e.~setting up and registering the given parameters) would always be expected to be 0.
Unfortunately, phases are not defined in the logs of Ardupilot-controlled drones, other than for a column called `MODE' which has three possible values: Stabilise, Auto and RTL (Return to Launch).

\begin{figure}[t]
    \centering
    \includegraphics[width=1\linewidth]{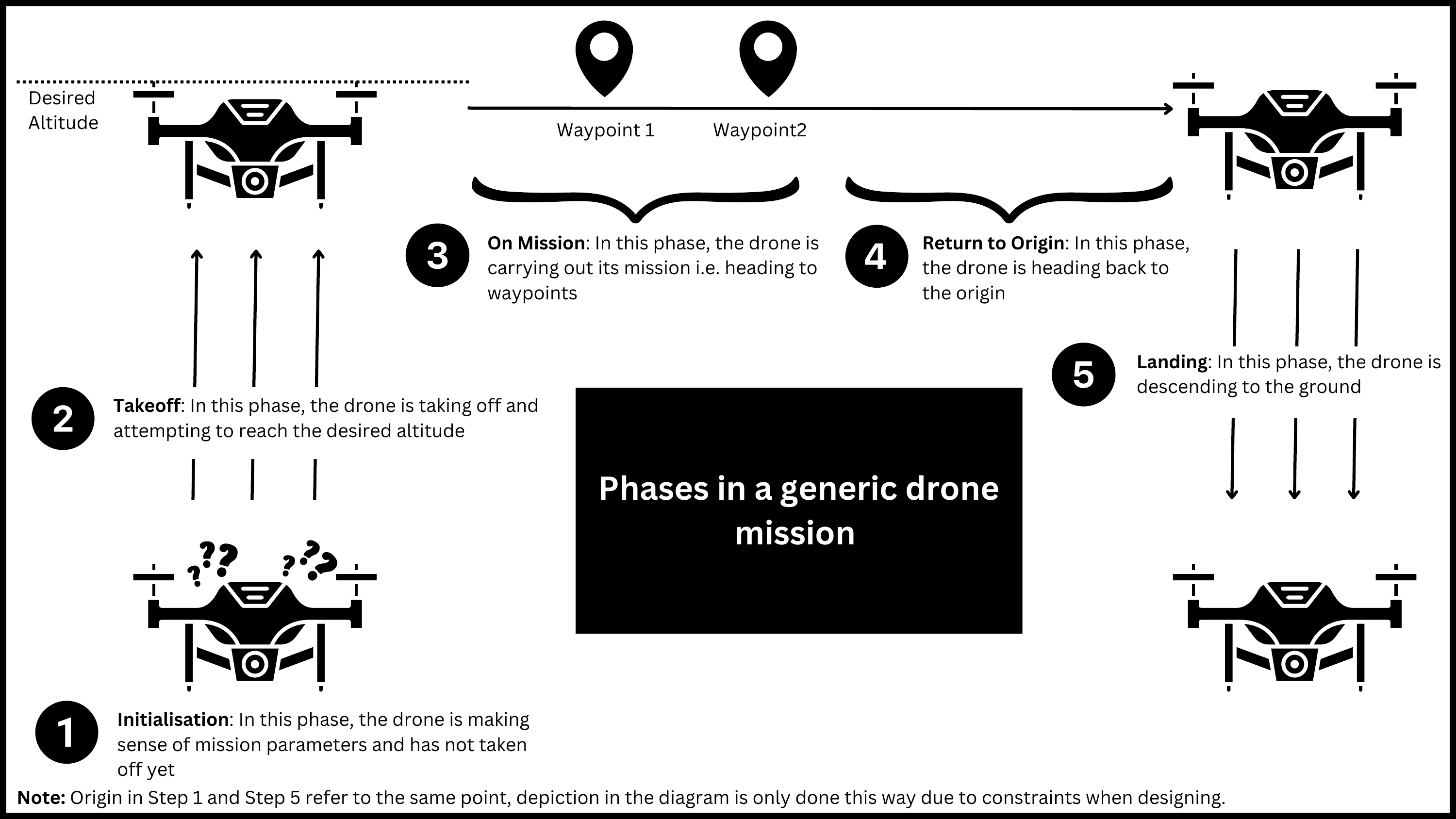}
    \caption{Phases in a generic drone mission}
    \label{fig:Phases}
    \end{figure}

Thus, we propose five phases of a generic drone mission that can be inferred from indicators in the logs.
After observing a multitude of missions, we identified five key phases that took place across all of them: Initialisation, Takeoff, On Mission, Return to Origin, and Landing.
A depiction of these phases can be seen in Figure~\ref{fig:Phases}.

In the \textbf{Initialisation} phase, there is a period of time in which the drone receives commands before taking off or starting a new mission from an airborne point.
During this time, the drone is essentially trying to make sense of the mission parameters and waypoints it was given.
Thus, we expect the same behaviour in this phase regardless of the particular mission, i.e.~no climbing, and no throttle, and thus an invariant range of values.

The \textbf{Takeoff} phase begins right after the mission parameters and configurations have been set.
The drone will attempt to take off from the starting position and reach the desired altitude of the mission.
Once the desired altitude is reached, the \textbf{On Mission} phase begins, in which the drone carries out most of its mission, i.e.~travelling via some set waypoints.

Upon completing the mission, the drone moves into the \textbf{Return to Origin} phase, in which it attempts to navigate back to its initial position, in the fastest and most linear way possible. Finally, once its original position has been reached, it will enter the \textbf{Landing} phase, in which it tries to safely descend to the ground.

Note that these phases are intended to be generic across different types of drones and missions, and we cannot assume that the current phase is given directly in the drone logs.
Rather, the current phase is derived according to some key indicators that are likely to be present.
For example, we use the relative altitude values to determine whether the drone is ready to transition from Takeoff to On Mission. The indicators for each phase can be seen in Table~\ref{fig:PhasesTable}.

\begin{table}[t]
	\centering
            \caption{Methodology used for deriving mission phases}
	    \label{fig:PhasesTable}
        \footnotesize
	\begin{tabular}{ll}
		\textbf{Phase} & \textbf{Conditions to check} \\ \hline
        INITIALISATION & ``MODE'' in original drone logs is STABILISE \\
        TAKEOFF & ``MODE'' has changed to AUTO \\
        ON MISSION & Desired takeoff altitude has been reached \\
        RETURN TO ORIGIN & Last waypoint has been reached \\
        LANDING & Original/Desired Coordinates have been
reached \\
	\end{tabular}
	%\vspace{-0.3cm}
\end{table}
%\vspace{-0.3cm}

\noindent
\textbf{Rule Construction.}
We implemented a form of association rule mining, the outcome of which can be transformed into rules that can be checked at runtime (i.e.~Boolean expressions over sensor and actuator values).
We classify these mined rules into two types: universal rules, which should hold at any point of the drone's mission, and phase-specific rules, which should hold throughout one of the five mission phases that were previously discussed.
Finally, we complement these with domain-specific rules based on domain expertise that may not be implicit from logs.

For CPS logs, specifically drones, rules usually involve a certain kind of range, due to the volatility of the values. We define these ranges based on the minimum and maximum values in our datasets. For example, during takeoff, the roll (rotation along front-back axis) value of the drone would be an abstract raw number like -0.00030842. After considering the data points from the relevant timestamps within a phase and processing it into a range, we would have a lower bound and upper bound. We then used the Apriori algorithm to find transactions that happened frequently during each phase---in our experiments, across a total of 30 datasets (Section~\ref{sec:evaluation}). The algorithm removes infrequent transactions, and uses a minimum support threshold to retain frequent transactions. These transactions are then translated into association rules which would eventually end up as our rules for checking.

\noindent\textbf{Universal Rules.}
Universal rules refer to the straightforward associations that the algorithm picked up without any changes to the input parameters.
These are rules that hold throughout the missions without violation.
As missions differ, there are not many of these rules, and they mostly had to do with sensors such as the GPS or Barometer.
Barring any faults, these would have constant values throughout a mission. 

\noindent\textbf{Phase-specific Rules.}
Phase-specific rules refer to those that the algorithm picked up for each of the specific phases only. These rules only hold throughout that particular phase. After mining the phase rules, we observed that there were some rules that might not necessarily hold true at all times in a phase due to possible instabilities. This does not mean that they should not be rules though. By setting a threshold of 99\%\ (100\%\ would be too rigid as there is the possibility of slight deviations due to the volatile nature of drones), we were able to find frequent relationships that held true except for certain minority data points. For example, in a scenario where the expected throttle is between 0 and 80, but one of the observations has a throttle value of 80.12. If not for this one data point, the range would hold. Then, we would examine the minority points, to see if they were negligible or explainable enough for us to still accept the rule or perhaps modify it. In the case of the throttle, we modified the rule to incorporate this value, as it was a negligible amount. 

\noindent\textbf{Domain-Specific Rules.}
{We discussed with a few domain experts from the drone industry and gathered some domain-specific rules, with respect to acceptable latency between the remote controller issuing a command and the drone acting upon it. For their specific use cases, drones are required to conduct long range missions, where drones may be affected by conditions like signal interference and obstacles like buildings and trees, which may result in delays in responding to the commands or signal lost. In such cases, since the drone may be visually out of sight, it is difficult to determine whether there is an anomaly or not. Even though the drone documentation may provide a specification of distances where a drone can safely operate in, it is often not clear what would be the acceptable latency. Therefore, we worked with the domain experts to conduct physical experiments with DJI drones and establish rules for determining anomalous latency issues. For example, from our experiments, we observed that the worst possible latency under normal circumstances is approximately 2 seconds for the DJI drones we tested and any latency more than 2 seconds should be considered an anomaly.

\noindent
\textbf{Unsupervised Learning.}
While rules can help us define boundaries to catch anomalies that we expect, there are bound to be anomalies that our rules are unable to catch.
This is where unsupervised learning comes in to fill the gap, allowing us to capture unexpected or unseen anomalies. Unsupervised learning also focuses on patterns rather than domain-knowledge, complementary to the rule-checking. We propose an ensemble of five models (K-Means, DBSCAN, OPTICS, LOF and SVM), to tap on their respective strengths to detect anomalies, by using a majority vote approach.

\noindent
\textbf{Identifying Anomalies by Voting.}
We then propose a voting ensemble based on the five models to finalise anomaly detection results, as each model has its own strengths in terms of anomaly detection. This is later confirmed in Section~\ref{sec:evaluation}, where we observe that $k$-means is effective in detecting anomalies from significantly more obvious anomalies, while SVM is effective for anomalies from datasets which are not too severe. We use the majority vote as a decision maker for whether a particular data point is deemed an anomaly or not. 

\noindent
\textbf{Integrating Rules and Unsupervised Models.}
We combine rule-based and unsupervised methods through a decision matrix, as shown in Table~\ref{decisionmatrix}. At runtime, both approaches independently check for anomalies. If an alert is triggered, the user receives a specific explanation---either identifying the violated rule or indicating that the unsupervised model has classified the point as anomalous. Within a sliding window, voting is performed on a per-timestamp basis, with each timestamp's result treated independently. In practice, based on the drone type, mission, and operator knowledge, alerts can be configured to trigger only when the proportion of anomalous timestamps within the window exceeds a specified threshold.

\begin{table}[!ht]
\caption{\label{decisionmatrix}Decision matrix for combining the two methods}
\footnotesize
\centering
\begin{tabular}{llll}
{\textbf{Rule-Based Method}}          & {\textbf{Clustering Ensemble Method}}             & {\textbf{Decision}} &    \\ \hline
\color{red}Rule(s) Broken    & \color{red}Majority Voted as Anomaly     & \color{red}Anomaly                 &    \\
\color{teal}{No Rule(s) Broken} & \color{red}Majority Voted as Anomaly     & \color{red}Anomaly                 &    \\
\color{red}Rule(s) Broken    & \color{teal}Majority Voted as Non-Anomaly & \color{red}Anomaly                 &    \\
\color{teal}No Rule(s) Broken & \color{teal}Majority Voted as Non-Anomaly & \color{teal}Non-Anomaly &   
\end{tabular}
\end{table}

\section{Evaluation}
\label{sec:experiment}
\label{sec:evaluation}

In this section, we present the design and results of our experiments to assess the efficacy of \coolname at detecting anomalies at runtime.
Our experiments are designed to answer the following Research Questions~(RQs):

\begin{itemize}[leftmargin=1.1em]
    \item \textbf{RQ1.} How effective is our approach at detecting anomalies in drone missions?

    \item \textbf{RQ2.} Is there a requirement for both parts of the approach? 

    \item \textbf{RQ3.} How does \coolname compare against the state-of-the-art?

    \item \textbf{RQ4.} Are the results produced by \coolname more interpretable?

    \item \textbf{RQ5.} Can \coolname be deployed at runtime?

\end{itemize}

\noindent
\subsection{Datasets Used}
Using the simulation environment mentioned in Section~\ref{sec:sim}, we generated representative datasets to train and validate our models, as well as datasets with anomalies for evaluation of our approach. The datasets generated were designed to cover a range of `normal' drone logs with no anomalies (which acted as the base of our training when evaluating our own validation datasets), as well as logs that are representative of several types of anomalies, ranging from harsh environmental conditions to different kinds of drone faults. Table \ref{tab:datasets} shows the statistics of each dataset.

\begin{table}[t]
\caption{Statistics of our datasets}
\label{tab:datasets}
    \centering
	{\small
	\begin{tabular}{lrr}
	  \hline
	 \textbf{Dataset} & \textbf{\#Entries} & \textbf{Anomalies?}  \\ \hline
	 Base & 450 & No Anomalies \\
	 Random & 278 & No Anomalies \\
	 Windy I & 674 & Anomalous Winds \\
	 Windy II & 424 & Anomalous Winds \\
	 Actuator & 470 & Faulty actuator \\
	 Sensor I (Roll) & 213 & Faulty sensor \\
	 Sensor II (Baro) & 356 & Faulty sensor \\\hline
	\end{tabular}
	} 
\end{table}

\noindent
\textbf{Base Dataset Generation.}
For the base datasets that we used in our models for training, we used the Ardupilot autopilot software, along with \textit{pymavlink} protocol and Gazebo simulator to run missions and collect log data from the drone missions. For the base dataset, we set four waypoints to complete a loop (simulating a real mission, using the same ones every time), and had the drone fly to each of them before returning to the origin.

\noindent
\textbf{Validation Datasets Generation.}
The simulations are configured differently to make sure each of the validation datasets was not identical to the base dataset. Some were created by injecting wind, engine faults, or manually injected sensor faults either throughout the whole mission or at certain periods throughout a mission. We then took note of when these faults were present, and labelled these timestamps as anomalies in these datasets. We then extracted \emph{only} the anomalous segments of the missions to have a 100\% anomalous dataset. We first created validation datasets of random missions, which were created in scenarios where we randomised the parameters such as the number of waypoints and mission coordinates, to ensure that they were significantly different from the datasets we used for training.

\noindent
\textbf{Simulating Harsh Environmental Conditions.}
We leveraged on the simulator's ability to simulate winds at user-desired speeds and directions. In cases of sensor or actuator faults, instability would be caused in the drone for various reasons. By applying a strong wind condition to the environment, we aimed to recreate this instability caused by these faults, to create a reflective simulation of them. These datasets had strong winds throughout the whole mission, to simulate sensor or actuator faults throughout. For these missions, there are severe instabilities throughout, and the drone did not manage to complete the mission by returning to the origin. We also created another dataset with respect to wind. As the strong wind that we simulated was too strong for the drone to complete the mission, we lowered the wind speeds minimally (at around 25\% of the strong wind's value) such that it could now do so. The instabilities caused in this simulation would still affect the drone heavily, with the only difference that the mission could be completed.

\noindent
\textbf{Simulating Actuator Faults.}
The simulator also had an option to cause faults in the drone's actuator. We could tweak a parameter, to have it only operating at a certain capacity. By setting it to lower values, the drone would not even be able to takeoff. We set the capacity to 0.7, which gave the drone enough to carry out the mission, but with many instabilities along the way. As the value was set throughout the mission, the drone had instabilities throughout as well, even during takeoff and landing.

\noindent
\textbf{Manual Injections to Simulate Sensor Failures.}
For these type of faults, we simulated sensor failures using manual injection. In many cases, when a sensor is faulty or malfunctioning, it will be `stuck', and will only display a certain value. A very common sensor fault involves values being stuck either at the minimum or maximum possible value. To simulate this, we chose a random mission and set one of the columns to its maximum possible value according to the simulator's documentation, and used the dataset as one with sensor failure. Specifically, we set the `roll' column to $\pi$, as that was the maximum possible value. We also created another dataset where we assumed the barometer was broken, showing 0 at all times. 

\noindent
\textbf{Hard Engine Failure Resulting in Crash.}
For the last type of fault, we wanted to simulate a drone during a crash. Similar to the actuator faults mentioned above, this time, instead of tweaking the capacity parameter by a little bit, we shut it off mid-flight to cause the drone to crash and not complete its mission. We then took the timestamps of when we injected the crash and extracted that part of the dataset for analysis.

\subsection{Results}

\noindent
\textbf{Effectiveness of Anomaly Detection (RQ1).} We applied \coolname to the logs within all of the aforementioned datasets, and present our results in Table~\ref{table:results1}.

\begin{table}[t] % put the table in here to make the section appear a bit longer.
\centering
\caption{Overall anomaly detection rates achieved by \coolname (\%). Random dataset contains no anomalies and therefore, we do not present \emph{Recall}. All other datasets are entirely anomalous and therefore, we do not present \emph{False Positives} or \emph{Precision}.}
\label{table:results1}
\footnotesize
\begin{tabular}{c c c} 
 %\hline
 & \textbf{Recall} & \textbf{False Positives} \\ [0.5ex] 
 \hline
 Random (No Anomalies) & N.A & 2.33\\ 
 %\hline
 Windy I (Mission not Completed) & 99.40 & N.A\\ 
 %\hline
 Windy II (Mission Completed) & 96.93 & N.A\\ 
 %\hline
 Actuator & 68.08 & N.A\\
 %\hline
 Sensor I (Roll) & 100.0 & N.A\\
 %\hline
 Sensor II (Baro) & 100.0 & N.A\\
 %\hline
 Crash & 98.64 & N.A\\  
 %\hline
\end{tabular}
\label{rules_anomaly-1}
\end{table}

For the various kinds of anomalies, our approach averaged an anomaly detection rate of 93.84\%, rising above 96\% for five of the faults.
For actuator faults, the rate is lower (68\%), which we believe is because the throttle loss induced was not severe enough to be detected as anomalous during all parts of the mission. As for the other faults, they were severe enough to be almost fully detected. Particularly for Windy I, where the wind speeds were so strong that the drone was unable to complete the mission based on the parameters and coordinates that were set. Due to the strong winds, the drone was unstable throughout the mission, and ended up crashing eventually. These occurrences were severe enough for RADD to almost fully detect them. In terms of false positives, \coolname kept this down to 2.33\%. We observed that the false positives were mostly isolated rows rather than sustained sequences, which means that they can actually be eliminated by having the system only alerting the operator when a sustained sequence of a few points (according to a threshold) are anomalous. Usually, these isolated rows could be due to the volatile nature of drones, where slight deviations or drifts can occur which causes the particular point to be slightly different from its neighbouring points.
These detection and false positive rates suggest that \coolname may be suitable and practical for anomaly detection in drones.

\noindent
\textbf{Ablation Study (RQ2).}
To prove the effectiveness and need for our approach to combine rules and unsupervised learning, we performed an ablation study, where we isolated the methods and ran the experiment with them solely, to find out the performance without the other method being involved. 

\noindent
\textbf{Rule-checking.} 
In terms of false positives, rule-checking alone fared well, with a false positive rate of 2.33\%. As for the various faults, it was able to detect anomalies effectively for five of the faults, with only 28.08\% for the Actuator dataset, as the throttle loss induced was not severe enough to trigger any breaking of the rules, which resulted in most of the points not being regarded as anomalies. More results can be seen in Table \ref{rules_anomaly-2}.

\begin{table}[t]
\centering
\caption{Anomaly detection rates of the rule-checking component only (\%)}
\label{rules_anomaly-2}
\footnotesize
\begin{tabular}{c c c} 
 %\hline
 & \textbf{Recall} & \textbf{False Positives}\\ [0.5ex] 
 \hline
 Random (No Anomalies) & N.A & 2.33 \\ 
 %\hline
 Windy I (Mission not Completed) & 99.40 & N.A\\ 
 %\hline
 Windy II (Mission Completed) & 96.93 & N.A\\ 
 %\hline
 Actuator & 28.08 & N.A \\
 %\hline
 Sensor I (Roll) & 100.0 & N.A \\
 %\hline
 Sensor II (Baro) & 100.0 & N.A \\
 %\hline
 Crash & 98.64 & N.A \\  
 %\hline
\end{tabular}
\end{table}

\noindent
\textbf{Unsupervised Learning.} 
Table \ref{clustering_anomaly} shows the results for the unsupervised learning portion of the approach. In terms of false positives, unsupervised learning alone fared very well, with a rate of only 0.58. Out of the five models, three had very low scores for false positives, and with a voting ensemble, this percentage was kept low. As for the anomaly detection rates for the various faults, the ensemble had a generally good performance for four of them, but had very low detection rates for Windy II and Sensor II specifically, which we elaborate on below.

\begin{table*}[t]
\centering
\caption{Anomaly detection rates of individual models and the ensemble (\%)}
\scriptsize
\vspace{-8pt}
\begin{tabular}{c c c c c c | c c} 
 %\hline
 \multicolumn{6}{c}{ } & \textbf{Ensemble} & \textbf{Ensemble} \\
 & K-Means & DBSCAN & OPTICS & LOF & SVM & \textbf{Recall} & \textbf{False Positives}\\ [0.5ex] 
 \hline
 Random (No Anomalies) & 0.0 & 0.58 & 0.0 & 87.13 & 15.2 & N.A & 0.58\\ 
 %\hline
 Windy I (Mission not Completed) & 55.34 & 3.1 & 76.56 & 97.03 & 84.42 & 75.07 & N.A\\ 
 %\hline
 Windy II (Mission Completed) & 0.0 & 3.1 & 50.0 & 96.11 & 1.65 & 1.17 & N.A\\ 
 %\hline
 Actuator & 59.78 & 0.85 & 74.89 & 99.57 & 75.1 & 64.25 & N.A\\
 %\hline
 Sensor I (Roll) & 100.0 & 2.83 & 75.94 & 100.0 & 100.0 & 100.0 & N.A\\
 %\hline
 Sensor II (Baro) & 0.0 & 0.84 & 24.43 & 69.66 & 3.65 & 2.52 & N.A\\
 %\hline
 Crash & 91.89 & 20.2 & 0.0 & 98.6 & 91.89 & 91.89 & N.A\\  
 %\hline
\end{tabular}

\label{clustering_anomaly}
\end{table*}

\noindent
\textbf{Windy I vs Windy II.}
In the scenario where we simulated heavy wind and the drone eventually crashed, the winds were so strong that the drone failed to complete its mission. Needless to say, these anomalies were caught almost fully by the rule-checking, and mostly by the unsupervised learning. We investigated and found that in the unsupervised learning check, the anomalies that were not caught belonged to a certain period, where the drone was trying to go the wind, and there were certain data points which were very similar in nature. Upon comparison, we found that they were similar to the normal `On Mission' data points, which could be why they were not flagged out.

As for the scenario where the wind was heavy, but not heavy enough to make the drone fail the mission, while rule-checking was able to flag out almost all the anomalies, unsupervised learning had a very low percentage. This could be because while there are instabilities now, they are constant instabilities, so the models are still able to cluster the points, and the points are not different enough to be recognised as anomalies.

A diagram depicting the two scenarios during takeoff can be seen in Figure~\ref{fig:windycomparison}.

\begin{figure}
\centering
\includegraphics[width=0.7\linewidth]{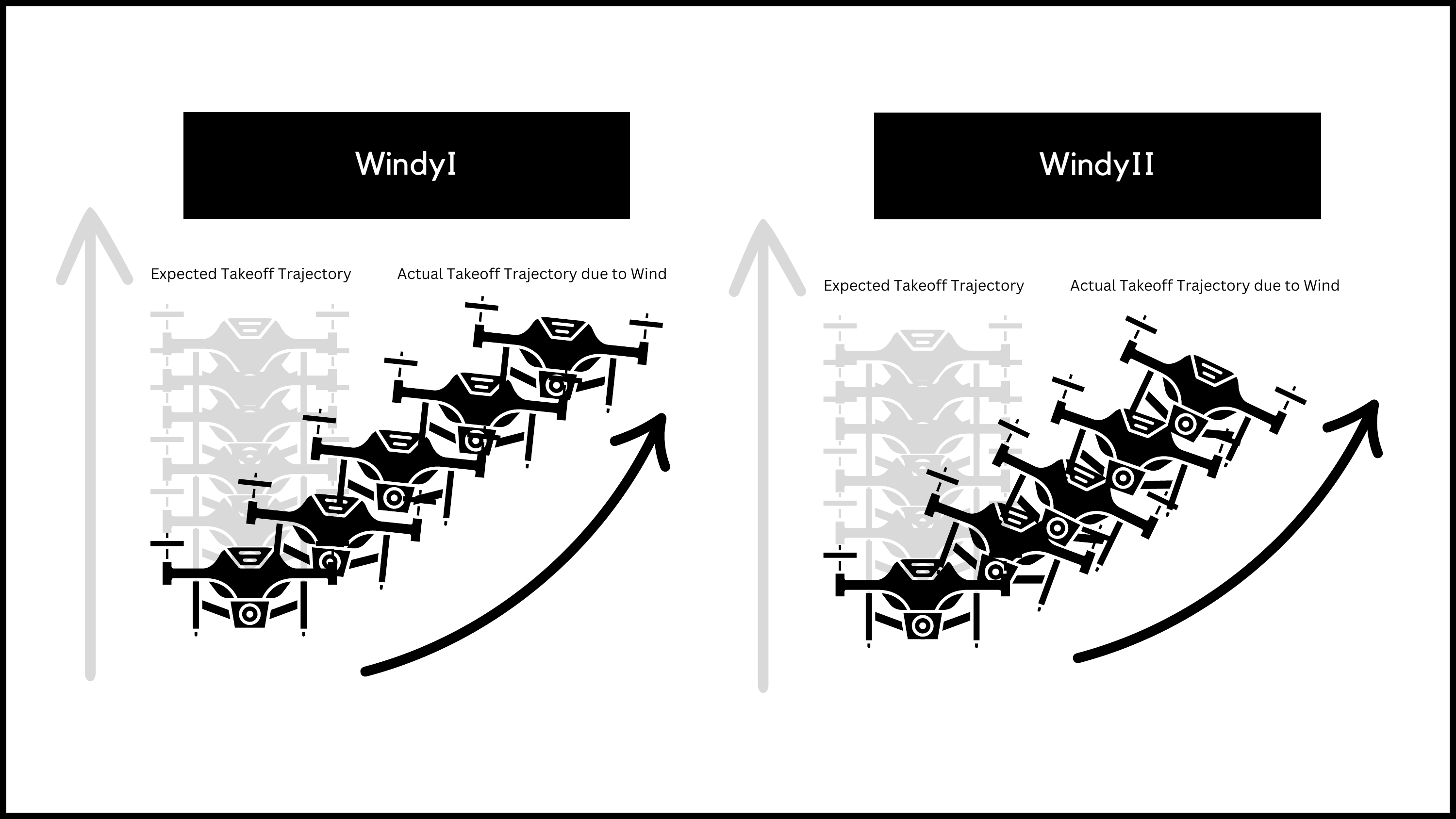}
\caption{Comparison between Windy I and Windy II}
\label{fig:windycomparison}
\vspace{-1.5em}
\end{figure}

\noindent
\textbf{Actuator.}
For the scenario where we simulated actuator faults, unsupervised learning performed a lot better than rule-checking. Upon investigation, we noticed that this is because to simulate the actuator faults, we adjusted their performance to 70\%. This caused slight instabilities, but not enough for rules to be broken. However, unsupervised learning was able to pick these instabilities up. This scenario shows an example where anomalies are only caught by unsupervised learning, and not by rule-checking.

\noindent
\textbf{Sensor (Roll).}
In this scenario, the Roll value was set to $\pi$. While it is a possible value according to the documentation, it is unlikely that the Roll of a drone will ever be that high. Thus, both our unsupervised learning and rule-checking methods detected it at full accuracy.

\noindent
\textbf{Sensor (Baro).}
In this scenario, the Barometer status remained at 0 throughout the entire mission. For the unsupervised learning model, this was not a significant anomaly, as the absence of sudden changes made it difficult to detect---the model primarily identifies deviations rather than constant incorrect values. However, our rule-based checking explicitly requires the Barometer status to be 1, allowing it to successfully detect the anomaly in this case.

\noindent
\textbf{Crash.}
For the crash scenario, as the drone was spiraling towards crashing, the instabilities were very obvious in most of the data points, explaining the high accuracy for both methods. 

\noindent
\textbf{Comparison to Baseline (RQ3).}
With respect to RQ3, we evaluated our approach with a baseline comparison against the LSTM model used in DronLomaly~\cite{Shar-et_al22a}. Using the same implementation of LSTM, we compared the performance with regards to our test datasets. Table \ref{table:7} shows the performance of the LSTM implementation, for 2, 2.5 and 3 standard deviations. Our approach fared better for all the type of faults in terms of detecting anomalies, and was able to achieve a lower false positive rate. In terms of cost, our approach also fared better than LSTM, due to the lesser training time required. The amount of data required for training was also significantly less than what was required for LSTM, with some of our base datasets only having about 500 data points. 

\begin{table}[!h]
\caption{Recall metrics for the DronLomaly LSTM detector}
\centering
\footnotesize
\begin{tabular}{c c c c c} 
 & $2\sigma$ & $2.5\sigma$ & $3\sigma$ & \coolname\\ [0.5ex] 
 \hline
 Random (No Anomalies) & 0.3784 & 0.1779 & 0.1301 & 0.0292\\ 
 %\hline
 Windy I (Mission not Completed) & 0.8352 & 0.0534 & 0.046 & 0.9940\\ 
 %\hline
  Windy II (Mission Completed) & 0.314 & 0.1362 & 0.0589 & 0.9693\\ 
 %\hline
 Actuator & 0.3662 & 0.3342 & 0.1572 & 0.6808\\ 
 %\hline
 Sensor I (Roll) & 1.0 & 1.0 & 1.0 & 1.0\\ 
 %\hline
 Sensor II (Baro) & 0.2643 & 0.1519 & 0.1038 & 1.0\\ 
 %\hline
  Crash & 0.9452 & 0.2577 & 0.0532 & 0.9864\\ 
 %\hline
 
\end{tabular}
\label{table:7}
\end{table}

\noindent
\textbf{Interpretability (RQ4).}
We designed a user study to evaluate the interpretability of our approach, i.e.~whether \coolname allows drone operators to identify what exactly is anomalous about the drone's behaviour.
We recruited 20 computer science undergraduates with varying levels of experience in drone operations and data analysis. The participants were first given a briefing on drones, their missions, and the various attributes and parameters that would be involved. We did not, however, show examples of what anomalous data looked like. This was to prevent them from forming ideas of what to look out for. Then, they were presented with anomalous drone missions generated through our simulations. They were then asked to assess and interpret the anomalies using both results: from rules generated by \coolname and from just raw data. Feedback indicated that our approach was much easier to analyse (averaging a 6.6 out of 7 on the Likert scale compared to the 3.2 scored by the raw data, with 7 being the easiest), with participants reporting a more intuitive understanding of the anomalies. In terms of actually identifying the anomalies, using our approach, 18 of the 20 participants were able to accurately determine what what the anomalous behaviour was. As for the raw data, only 3 participants were able to identify the anomaly. This was achieved through the transparent presentation of rules broken, which allowed users to identify anomalous behaviors more quickly and confidently. In contrast, the raw data often required more time and led to uncertainty in decision-making. The study highlights the potential of our approach to enhance operational workflows by improving anomaly detection accuracy, especially in critical situations where swift identification of issues is essential to maintaining drone safety and performance. More details of the study can be found online\footnote{\url{https://sites.google.com/view/raddstudy/home}}.

\noindent
\textbf{Deploying at Runtime (RQ5).}
To test runtime deployment of our approach, we deployed it on a Raspberry Pi 4 Model B with 8 GB memory, which can actually be used to control/fly a drone. On the device, we then ran the approach and predicted each observation entry in the random validation dataset, one at a time to simulate an actual log pipeline feeding it data. As the predictions were being made one at a time, we collected the time taken for each prediction for analysis. The results for K-Means, DBSCAN, LOF and SVM can be seen in Figure~\ref{fig:boxplot1}. The results for OPTICS can be seen in Figure~\ref{fig:boxplot2}. We can see that $k$-means, DBSCAN, and LOF are very lightweight, taking about 10 milliseconds (0.01 seconds) for each prediction on average, whereas LOF is a little bit heavier, taking about 20 milliseconds (0.02 seconds) for each prediction on average. OPTICS, on the other hand, requires more time as it has to re-fit the training data along with the new data to make a prediction. This makes the average timing significantly higher than the rest, at about 1250 milliseconds (1.25 seconds). As for the rule checking, the time in milliseconds was almost negligible. Adding all the timings together, it would take about 1.3 seconds to analyse each data point. Taking into account the amount of time it takes for a drone to actually crash to the ground as well as how long it would take for the user to read an alert and make a decision, this timing is reasonable for deployment at runtime.

\begin{figure}[t]
\centering
\includegraphics[width=0.8\linewidth]{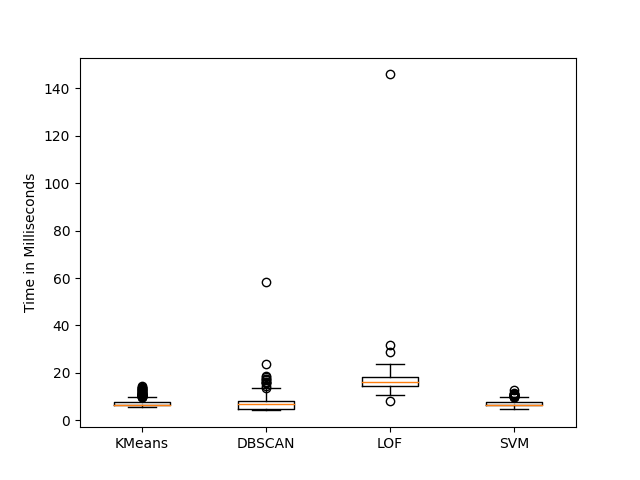}
\caption{Boxplot of prediction times for unsupervised models}
\label{fig:boxplot1}
\end{figure}

\begin{figure}[t]
\centering
\includegraphics[width=0.8\linewidth]{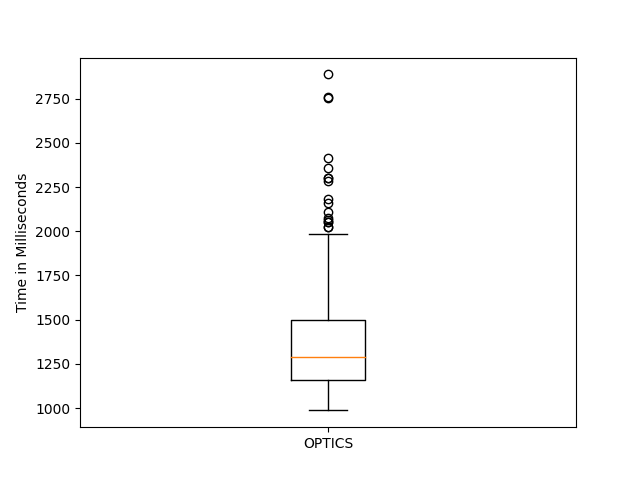}
\caption{Boxplot of Prediction Times for OPTICS on Raspberry Pi}
\vspace{-2em}
\label{fig:boxplot2}
\end{figure}

\noindent
\textbf{Threats to Validity.}
\label{sec:threats}
\textit{(1)~Lack of variety in simulator-generated datasets}. As the simulator has limited environment variables for us to experiment with, the simulation data is not comprehensive enough to cover all possible real scenarios. Many faults—such as those caused by human error, which can occur in countless unpredictable ways—remain beyond our reach. As a result, while we can confidently say that our simulations test certain aspects, we cannot generalise the models to all possible fault conditions in drones. A potential mitigation strategy for future work is to incorporate real-world flight log data and adapt it to create more realistic, domain-specific scenarios.

\smallskip
\noindent
\textit{(2)~Running the analyses on raw data}. We chose not to standardise the data because the selected features—representing the drone’s position, speed, and angles—are critical absolute values that are inherently sensitive to changes. Applying a scaler could have distorted these values and compromised their true meaning, potentially affecting the accuracy of anomaly detection.

\section{Related Work}
\label{sec:related-work}

Our study is closely related to anomaly detection for CPSs, which typically involves profiling behaviours or patterns from observations such as sensor readings and actuator outputs, and identifying those that differ significantly from the expected behaviour. 
In general, these approaches can be broadly categorised into rule-based approaches, supervised learning-based approaches, and unsupervised learning-based approaches. 

\noindent
\textbf{Log-Based Defence Mechanisms for CPSs.} Anomalies can be detected by evaluating data points against rules (or invariants), in which expected relationships between sensors and actuators are extracted from data sets~\cite{Feng-et_al19a}, control programs and the laws of physics~\cite{Choi-et_al18a}, or alternatively, are defined based on \emph{a priori} knowledge~\cite{Adepu-Mathur16a,Yoong-et_al21a}.
To our knowledge, rules for detecting anomalies in drone systems have not been explored. Therefore, in this work, we define some domain specific rules for drones and also propose a rule mining algorithm on the drone logs to automatically derive a specific rule set.

If labelled data is available, then supervised learning approaches like SVM can be used to train a model that classifies new data points as normal or anomalous~\cite{Chen-Poskitt-Sun18a}. However, it may be hard to gather labeled anomaly data, especially since anomalies are often unknown or unpredictable for complex systems like CPSs. In this case, unsupervised approaches like clustering~\cite{Kiss-et_al15a}, local outlier factor~\cite{7925416}, and one-class SVM~\cite{Inoue-et_al17a} can be used to learn a decision boundary that encapsulates the normal data points.
\emph{Clustering} approaches have been reviewed by Aghabozorgi et. al.~\cite{aghabozorgi2015time} and Alqahtani et. al.~\cite{alqahtani2021deep}. According to these surveys, there is a considerable focus on using clustering to detect patterns in CPSs, for example, for detecting spatial patterns in Silicon wafers~\cite{liukkonen2018recognition}, for detecting machine wear and predictive maintenance~\cite{amruthnath2018research}, and for detecting anomalies in traffic data~\cite{munz2007traffic}. Outlier detection algorithms such as One-Class SVM~\cite{manevitz2001one} and Local Outlier Factor (LOF)~\cite{ma2016density} have also been applied for anomaly detection on CPS log data and achieved promising results. But, to our knowledge, these approaches have not been applied in the context of anomaly detection for drone systems. A recent work \cite{silalahi2024severity} also focuses on the severity being a key factor in separating critical anomalies and less severe ones. Drone anomalies might not always be critical or severe, so this is an important motivation for us to consider as well.

Deep learning approaches, in which neural networks learn the complex non-linear patterns in data sets, can be used as time series predictors, with the anomaly alarm raised when predicted physical states differ significantly from the ones eventually observed. For example, in \cite{feng2017multi, Goh-et_al17a, Kravchik-Shabtai18a, Kravchik-Shabtai22a}, LSTM, RNN, and CNN were used to detect anomalies or cyber attacks to industry control systems. 
DronLomaly~\cite{Shar-et_al22a} has taken some first steps towards applying anomaly detection techniques to drones, specifically, by utilising LSTM netural networks to predict future sensor/actuator states and marking anomalies when the actually observed states diverge. However, DronLomaly has the risk of overfitting to the training data from sample flying logs as its model is trained from a specific mission log and is used to detect anomalies when the drone is conducting the same mission. Models must be able to \emph{generalise}, as no two missions are the same: flying in an open field is different to flying in a crowded city, not to mention the differences that arise due to weather.
Generally, deep learning approaches lack \emph{interpretability}. In the context of drone missions, this makes it difficult for drone operators to understand the nature of an anomaly and decide what appropriate action to take, in time.
Furthermore, drones are often required to satisfy properties that are simple to define with \emph{domain knowledge} such as acceptable latency, but may not always be easily learnt from log data. These findings motivated us to explore alternative anomaly detection approaches that overcome the specific challenges for drones.

\noindent
\textbf{Defence and Analysis Techniques for Drones.} In \cite{renduchintala2017drone, mekala2019digital, kumar2021drone}, forensics frameworks for examining the drone's activities via its logs, upon the occurrence of an incident, have been presented. These approaches deal with after-the-fact drone forensics analyses, e.g.~to find evidence for the court of law.
Drone-specific fuzzing approaches such as PGFuzz~\cite{kim:pgfuzz} and LGDFuzzer~\cite{lgdfuzzer} have been proposed to detect bugs in drone control programs. These approaches execute a given drone mission with perturbed drone control parameters and detect if the drone violates safety constraints such as crashes. Such approaches are useful at detecting input validation and semantic bugs in drone control programs but are not designed for detecting anomalies such as sensor/actuator failures during runtime. 

There are also approaches for detecting cyber attacks on drones at runtime. For example, \cite{gudla2018defense} describes a Moving Target Defense approach focusing on communication network security and intrusion detection, \cite{zhang2019securing} proposes algorithms at the network layer to secure wireless channels used in drones to ground communication in 5G networks, and \cite{condomines2019network} discusses anomaly detection techniques focusing on distributed denial of service attacks in drone networks. \cite{wang2024exploratory} talked about compilating the features and patterns of anomalies in drone logs.  Our work in this paper on log-based runtime detection of anomalies in drones can complement such existing defence approaches which have been focusing more on the network and communication layer of drone systems.

\section{Conclusion and Future Work}
\label{sec:conclusion}
\label{sec:future-work}

This paper presents \coolname, an anomaly detection approach for drones based on the integration of rule-checking and unsupervised learning.
The combination of these methods allows \coolname to generalise to a diversity of weather conditions and scenarios, while ensuring that detected anomalies are also more interpretable than those caught by deep neural networks.
We have implemented \coolname for the ArduPilot drone software in the Gazebo simulator, mining 44 rules across five derived mission phases, and training an ensemble of five unsupervised models.
We have evaluated \coolname with various drone log datasets and found that it is able to detect an average of 93.84\% of anomalies over six types of faults (rising to 99\% for heavy wind and sensor faults) with a low false positive rate (2.33\%), and exceed the performance of a widely-used LSTM-based detector for drones.
Finally, in an ablation study we have observed that the integration of approaches exceeds the performance of any one part, and that they are efficient enough to be deployed as runtime checkers.

There are a number of interesting directions of future work.
First, we would like to explore whether the ensemble can be improved, e.g.~by exploring weighted voting mechanisms based on the level of `trust' in individual models. Currently, only a basic voting system is used, which can be improved to give us more information about the anomaly if weights were involved.
Second, we would like to extend the study to datasets in which multiple types of anomalies are happening at once (or in an interleaved manner).
Third, we want to explore incremental rule updates during missions and their phases to adapt to unforeseen scenarios.
Finally, we would be interested in exploring strategies for automatically recovering from anomalies once they have been identified by \coolname.

\bibliography{references}
\bibliographystyle{splncs04}
\end{document}